\documentclass[draft]{birkjour}
\usepackage[noadjust]{cite}
\usepackage{xcolor}
\RequirePackage[all]{xy}

\usepackage{hyperref}

\newtheorem{thm}{Theorem}[section]

\newtheorem{prop}[thm]{Proposition}
\theoremstyle{definition}
\newtheorem{defn}[thm]{Definition}
\theoremstyle{remark}

\newtheorem*{ex}{Example}
\numberwithin{equation}{section}

\newcommand{\BibTeX}{B\kern-0.1emi\kern-0.017emb\kern-0.15em\TeX}
\newcommand{\XYpic}{$\mathrm{X\kern-0.3em\raisebox{-0.18em}{Y}}$-$\mathrm{pic}\,$}

\newcommand{\cl}{C \kern -0.1em \ell}  

\newcommand{\bx}{\mathbf{x}}
\newcommand{\by}{\mathbf{y}}
\newcommand{\ba}{\mathbf{a}}
\newcommand{\bb}{\mathbf{b}}

\newcommand{\BZ}{\mathbb{Z}}
\newcommand{\BR}{\mathbb{R}}

\newcommand{\ed}{\end{document}}
\begin{document}

%
%
%
%
%
%
%
%
%



\newcommand{\rd}{\mathrm{d}}
\def\g{\mathfrak{g}}
\def\Lieg{{\mathrm{Lie}(G)}}
\def\gr{\mathrm{g}}
\def\h{\mathfrak{h}}

\newcommand*{\R}{{\mathbb R}}
\def\tr{{\it tr\/}}

\newcommand{\namecite}{\small\color{magenta}}

\newcommand{\calA}{\mathcal{A}}
\newcommand{\calB}{\mathcal{B}}
\newcommand{\calC}{\mathcal{C}}
\newcommand{\calD}{\mathcal{D}}
\newcommand{\calE}{\mathcal{E}}
\newcommand{\calG}{\mathcal{G}}
\newcommand{\calH}{\mathcal{H}}
\newcommand{\calI}{\mathcal{I}}
\newcommand{\calJ}{\mathcal{J}}
\newcommand{\calL}{\mathcal{L}}
\newcommand{\calM}{\mathcal{M}}
\newcommand{\calN}{\mathcal{N}}
\newcommand{\calX}{\mathcal{X}}
\newcommand{\calO}{\mathcal{O}}
\newcommand{\calR}{\mathcal{R}}
\newcommand{\calS}{\mathcal{S}}
\newcommand{\calV}{\mathcal{V}}

\newcommand{\cev}[1]{{\stackrel{\leftarrow}{#1}}}

\def\delnl{\delta_{\hbox{\sc NL}}}
\def\ep{\epsilon}

\def\delzero{\delta_0}
\def\brs{\delta}
\def\Somark{.\raisebox{1ex}{.}.}
\def\Bemark{\raisebox{1ex}{.}.\raisebox{1ex}{.}}

\def\delh{{\delta_1}}
\def\dela{{\delta_2}}
\def\delb{{\delta_3}}
\def\eph{{\epsilon_1}}
\def\epa{{\epsilon_2}}
\def\epb{{\epsilon_3}}
\def\dM{{d_M}}

\newcommand\ad{{\rm ad}}
\newcommand{\rdif}{\overleftarrow{\partial}} 

\newcommand\gh{{\rm gh}}
\newcommand\negative{{\rm neg}}
\newcommand{\lb}[2]{[\![#1\,,#2]\!]}
\newcommand{\ld}{\overrightarrow{\partial}} 
\newcommand{\sbv}[2]{{\{{{#1},{#2}}\}}}
\newcommand{\sbvike}[2]{{\{{{#1},{#2}}\}}}
\newcommand{\tbv}[2]{{\{{{#1},{#2}}\}_{\calL}}}
\newcommand{\abv}[2]{{[{{#1},{#2}}]_{(\Theta,\alpha)}}}
\newcommand{\courant}[2]{{[{{#1},{#2}}]_D}}
\newcommand{\couranth}[2]{{[{{#1},{#2}}]_H}}
\newcommand{\courantr}[2]{{[{{#1},{#2}}]_R^{\pi}}}
\newcommand{\schouten}[2]{{[{{#1},{#2}}]_S}}

\newcommand{\bracket}[2]{\langle #1,\,#2\rangle}
\newcommand{\bbracket}[2]{\Biggl\langle #1,\,#2\Biggr\rangle}
\newcommand{\inner}[2]{{({{#1},{#2}})}}
\newcommand{\inwedbra}[2]{\langle #1\inwedge\,#2\rangle}

\newcommand{\bsbv}[2]{{\{{{#1},{#2}}\}_{b}}}
\newcommand{\ssbv}[2]{{\{{{#1},{#2}}\}}}
\newcommand{\lbv}[2]{{\{{{#1},{#2}}\}_{\calL}}}

\def\bold#1{#1\l[lap{$#1$\hskip.3pt}\llap{$#1$\hskip.4pt}\llap{$#1$\hskip.5pt}}
\def\bbold#1{\bold{\bold#1}}
\def\sbold#1{#1\llap{$#1$\hskip.1pt}\llap{$#1$\hskip.2pt}\llap{$#1$\hskip.3pt}}
\def\sbbold#1{\sbold{\sbold#1}}
\def\bos#1{\bbold#1}

\def\bF{\mbox{\boldmath $F$}}
\def\bG{\mbox{\boldmath $G$}}
\def\bH{\mbox{\boldmath $H$}}
\def\bV{\mbox{\boldmath $V$}}
\def\bX{\mbox{\boldmath $X$}}
\def\bY{\mbox{\boldmath $Y$}}
\def\bZ{\mbox{\boldmath $Z$}}

\def\bx{\mbox{\boldmath $X$}}
\def\bxi{\mbox{$\xi$}}
\def\bzeta{\mbox{$\zeta$}}
\def\bq{\mbox{\boldmath $q$}}
\def\bp{\mbox{\boldmath $p$}}
\def\bbx{\mbox{\boldmath $x$}}
\def\bbxi{\mbox{\boldmath $\xi$}}
\def\bbq{\mbox{\boldmath $q$}}
\def\bbp{\mbox{\boldmath $p$}}
\def\bbe{\mbox{\boldmath $e$}}
\def\bbc{\mbox{\boldmath $c$}}
\def\bbt{\mbox{\boldmath $t$}}
\def\bbF{\mbox{\boldmath $F$}}
\def\bbu{\mbox{\boldmath $u$}}
\def\bbv{\mbox{\boldmath $v$}}
\def\bbw{\mbox{\boldmath $w$}}
\def\bbz{\mbox{\boldmath $z$}}
\def\brho{\mbox{\boldmath $\rho$}}

\newcommand{\floor}[1]{{\lfloor #1 \rfloor}}

\def\ba{\mbox{\boldmath $A$}}
\def\bb{\mbox{\boldmath $B$}}
\def\by{\mbox{\boldmath $Y$}}
\def\bz{\mbox{\boldmath $Z$}}
\def\bR{\mbox{\boldmath $R$}}
\def\bT{\mbox{\boldmath $T$}}
\def\bC{\mbox{\boldmath $C$}}
\def\bPhi{\mbox{\boldmath $\Phi$}}
\def\bh{\mbox{\boldmath $h$}}
\def\bphi{\mbox{\boldmath $\phi$}}
\def\bPsi{\mbox{\boldmath $\Psi$}}
\def\bvartheta{\mbox{\boldmath $\vartheta$}}
\def\bomega{\mbox{\boldmath $\omega$}}
\def\bTheta{\mbox{$S_1$}}
\def\bQ{\mbox{\boldmath $Q$}}

\def\bDelta{\mbox{\boldmath $\Delta$}}
\def\bdelta{\mbox{\boldmath $\delta$}}
\def\balpha{\mbox{\boldmath $\alpha$}}
\def\bpsi{\mbox{\boldmath $\psi$}}
\def\bchi{\mbox{\boldmath $\chi$}}

\def\lcf{\lbrack\! \lbrack}
\def\rcf{\rbrack\! \rbrack}

\newcommand{\dbv}{{\Delta_{\hbox{\sc BV}}}}
\newcommand{\OmBV}{{\Omega_{\hbox{\sc BV}}}}
\newcommand{\bOmBV}{\mbox{\boldmath ${\Omega_{\hbox{\sc BV}}}$}}

\newcommand{\Sprime}{{\Sigma}}

\newtheorem{Example}{Example}
\newcommand{\Map}{{\rm Map}}
\newcommand{\ev}{{\rm ev}}

\newcommand{\nom}{\nonumber}
\newcommand{\beq}{\begin{equation}}
\newcommand{\eeq}{\end{equation}}
\newcommand{\bea}{\begin{eqnarray*}}
\newcommand{\eea}{\end{eqnarray*}}
\newcommand{\beqa}{\begin{eqnarray}}
\newcommand{\eeqa}{\end{eqnarray}}
\newcommand{\vph}{\boldsymbol{\varphi}}
\newcommand{\vsi}{\boldsymbol{\sigma}}
\newcommand{\btheta}{\boldsymbol{\theta}}

\newcommand{\bJ}{\boldsymbol{J}}
\newcommand{\vrh}{\boldsymbol{\rho}}
\newcommand{\val}{{\boldsymbol{\alpha}}}
\newcommand{\vbe}{\boldsymbol{\beta}}
\newcommand{\vaa}{\boldsymbol{a}}
\newcommand{\vab}{\boldsymbol{b}}
\newcommand{\vom}{\boldsymbol{\omega}}
\newcommand{\veta}{\boldsymbol{\eta}}
\newcommand{\vxi}{\boldsymbol{\xi}}
\newcommand{\vchi}{{\boldsymbol{\chi}}}
\newcommand{\del}{\partial}
\newcommand{\ti}{\tilde}
\newcommand{\olv}{\overleftarrow}
\newcommand{\orv}{\overrightarrow}

\def\bOmega{\mbox{\boldmath $\omega$}}
\def\bOmega{\mbox{$\Omega$}}
\newcommand{\pbv}[2]{{\{{{#1},{#2}}\}}}

\newcommand{\cfun}[1]{\langle #1 \rangle}
\def\bvarphi{\mbox{\boldmath $\varphi$}}

\newcommand{\hatj}{\widehat{j}}
\newcommand{\tildej}{\widetilde{j}}
\def\bcalC{\mbox{\boldmath $\calC$}}
\newcommand{\gcourant}[2]{{[{{#1},{#2}}]}}
\newcommand{\be}{\boldsymbol{e}}

\newcommand{\cJ}{{J_{cl}}}
\newcommand{\cepsilon}{{\epsilon_{cl}}}

\newcommand{\liethree}[2]{{[{{#1},{#2}}]_3}}
\newcommand{\proj}{{pr}}

\newcommand\qone{{\eta}}
\def\bqone{\mbox{\boldmath $\eta$}}

\newcommand\eeqref{\eqref}

\newcommand{\ty}{\tilde{y}}

\newcommand{\hQ}{Q}

\makeatletter
\newcommand\xleftrightarrow[2][]{%
  \ext@arrow 9999{\longleftrightarrowfill@}{#1}{#2}}
\newcommand\longleftrightarrowfill@{%
  \arrowfill@\leftarrow\relbar\rightarrow}
\makeatother

\newcommand{\cp}{{\check{p}}}
\newcommand{\cq}{{\check{q}}}
\newcommand{\cx}{{\check{x}}}
\newcommand{\cxi}{{\check{\xi}}}
\newcommand{\hx}{{\hat{x}}}
\newcommand{\hxi}{{\hat{\xi}}}
\newcommand{\tB}{{\tilde{B}}}
\newcommand{\tbeta}{{\tilde{\beta}}}
\newcommand{\hbeta}{{\hat{\beta}}}
\newcommand{\hB}{{\hat{B}}}
\newcommand{\hq}{{\hat{q}}}
\newcommand{\hp}{{\hat{p}}}
\newcommand{\hj}{{\widehat{j}}}
\newcommand{\hcM}{{\widehat{\mathcal{M}}}}
\newcommand{\hM}{{\widehat{M}}}
\newcommand{\cQ}{{\mathcal{Q}}}
\newcommand{\cM}{{\mathcal{M}}}
\newcommand{\tM}{{\tilde{M}}}
\newcommand{\tq}{{\tilde{q}}}
\newcommand{\tp}{{\tilde{p}}}
\newcommand{\txi}{{\tilde{\xi}}}
\newcommand{\tx}{{\tilde{x}}}
\newcommand{\tQ}{{\tilde{Q}}}
\newcommand{\tP}{{\tilde{P}}}
\newcommand{\tXi}{{\tilde{\Xi}}}
\newcommand{\tX}{{\tilde{X}}}
\newcommand{\teta}{{\tilde{\eta}}}
\newcommand{\tpi}{{\tilde{\pi}}}
\newcommand{\tcM}{{\tilde{\cM}}}
\newcommand{\tj}{{\tilde{j}}}
\newcommand{\Si}{{\Sigma}}
\newcommand{\si}{{\sigma}}
\newcommand{\tSi}{{\tilde{\Sigma}}}
\newcommand{\tsi}{{\tilde{\sigma}}}
\newcommand{\tpartial}{{\tilde{\partial}}}
\newcommand{\td}{{\tilde{d}}}
\newcommand{\te}{{\tilde{e}}}
\newcommand{\tu}{{\tilde{u}}}
\newcommand{\tpar}{{\tilde{\partial}}}
\newcommand{\Sp}{{\;\,}}
\newcommand{\tR}{{\tilde{R}}}
\newcommand{\mT}{{\mathfrak{T}}}
\newcommand{\mbT}{{\bar{\mathfrak{T}}}}
\newcommand{\tstar}{{\tilde{\star}}}
\def\px#1{\frac{\partial}{\partial x^{#1}}}
\def\ptx#1{\frac{\partial}{\partial \tx_{#1}}}
\def\pp#1{\frac{\partial}{\partial p_{#1}}}
\def\ptp#1{\frac{\partial}{\partial \tp^{#1}}}

\newcommand{\wcM}{{\widehat{\cM}}}
\newcommand{\wM}{{\widehat{M}}}

\newcommand{\Det}{\mathrm{Det}}

\newcommand{\calMC}{\mathcal{MC}}

\newcommand{\Hom}{{\rm Hom}}
\newcommand{\md}{{d}}

\def\baa{\mbox{\boldmath $A^{\prime}$}}
\def\baaa{\mbox{\boldmath $A^{\prime\prime}$}}
\def\pii{\mbox{$\pi^{\prime}{}$}}
\def\BB{\mbox{$B$}}
\def\BBB{\mbox{$B^{\prime}$}}

\def\tpi{\tilde{\pi}{}}
\def\beqa{\begin{eqnarray}}
\def\eeqa{\end{eqnarray}}
\def\beq{\begin{equation}}
\def\eeq{\end{equation}}
\def\be{\begin{equation}}
\def\ee{\end{equation}}

\def\xzero{X{}}
\def\tilx{\widetilde{X}{}}
\def\tila{\widetilde{A}{}}
\def\tilz{\widetilde{z}{}}

\def\bbd{{{\rd}_{\calX}}}
\def\sd{\mbox{\boldmath $\mathrm{d}$}}

\newcommand{\Gammaz}{\stackrel{\circ}{\Gamma}}
\newcommand{\nablaz}{\stackrel{\circ}{\nabla}}
\newcommand{\Az}{\stackrel{\circ}{A}}
\newcommand{\tilaz}{\stackrel{\circ}{\widetilde{A}^{\nabla}}}

\newcommand{\uX}{{{X}}{}}
\newcommand{\uAplus}{{\underline{A}^+}{}}
\newcommand{\ucplus}{{\underline{c}^+}{}}
\newcommand{\uc}{{\underline{c}}{}}
\newcommand{\uA}{{\underline{A}}{}}
\newcommand{\uXplus}{{\underline{X}^+}{}}
\newcommand{\uphi}{{\underline{\phi}}{}}
\newcommand{\upsi}{{\underline{\psi}}{}}
\newcommand{\uxi}{{\underline{\xi}}{}}

\newcommand{\udx}{{\underline{\rd x}}{}}
\newcommand{\upartial}{{\underline{\partial}}{}}

\newcommand{\uu}{{\underline{u}}{}}
\newcommand{\uv}{{\underline{v}}{}}
\newcommand{\ue}{{\underline{e}}{}}
\newcommand{\ualpha}{{\underline{\alpha}}{}}
\newcommand{\ubeta}{{\underline{\beta}}{}}
\newcommand{\uJ}{{\underline{J}}{}}

\newcommand{\momega}{{\omega}}
\newcommand{\gomega}{{\omega}_{grad}}
\newcommand{\tmu}{{\tilde{\mu}}}
\newcommand{\hmu}{{\widehat{\mu}}}
\newcommand{\tH}{{\tilde{H}}}

\newcommand{\rde}{{{}^E \rd}}
\newcommand{\rddr}{{\rd_{\mathrm{dR}}}}

\def\hrho{\widehat{\rho}{}}

\newcommand{\rhoe}{\rho}
\newcommand{\rhoa}{{\rho}}

\newcommand{\msomega}{{\omega}}

\def\sd{\mbox{\boldmath $\mathrm{d}$}}

\def\tbx{{\widetilde{\bx}}}
\newcommand{\anabla}{{{}^E \nabla}}
\newcommand{\baS}{{{}^E S}}
\newcommand{\tS}{{{S_{BV}}}}


\def\a{\alpha} \def\b{\beta} \def\g{\gamma} \def\G{\Gamma} \def\d{\delta} \def\D{\Delta}
\def\e{\epsilon} \def\vare{\varepsilon}
\def\f{\phi} \def\F{\Phi} \def\h{\eta} \def\k{\kappa}
\def\r{\rho}


\def\R{{\mathbb R}} \def\C{{\mathbb C}} \def\N{{\mathbb N}} \def\X{\mathbb X} \def \A{\mathbb A}
\def \F{\mathbb F} \def \B{\mathbb B}
\def \L{\mathbb L} \def\Z{{\mathbb Z}} 

\newcommand{\DD}{{\mathrm{D}}}

\def\rc#1{\textcolor{red}{#1}}
\def\bc#1{\textcolor{blue}{#1}}
\def\mc#1{\textcolor{magenta}{#1}}


\title[]
 {Q-Manifolds and Sigma Models}
\author[]{Noriaki Ikeda}
%
\address{%
Department of Mathematical Sciences,
Ritsumeikan University \\
Kusatsu, Shiga 525-8577, Japan
}
\email{nikeda@se.ritsumei.ac.jp}
\thanks{}
\subjclass{
}
\date{\today}
\dedicatory{Last Revised:\\ \today}
\begin{abstract}
Recent developments of Batalin-Vilkovisky (BV) formalism and related geometry 
are reviewed. 
Mathematical structures of BV formalism are summarized as a Q-manifold 
and a QP-manifold.
Lie algebras, Lie algebroids and other higher algebroids 
are explained as typical examples of Q- and QP-manifolds.
Finally, the BV action functionals are constructed by geometric structures of Lie algebroids and Courant algebroids.
\end{abstract}
\label{page:firstblob}
\maketitle
\tableofcontents

\section{Introduction}

Batalin-Vilkovisky (BV) formalism of gauge theories 
is a canonical formulation for quantization of 
a classical field theory with gauge symmetries.
In the BRST-BV quantization in a gauge theory, the so-called ghost fields which 
are odd or even variables, are introduced. 


Graded manifolds, especially, Q-manifolds and QP-manifolds are mathematical 
basis of BV formalism. 

A BV action functional $S_{BV}$ is a homological function 
generating BRST transformations and carries all the physical information.
However, the construction of the BV action functional $S_{BV}$
is not easy. 
Especially, geometrical interpretation of $S_{BV}$ is not well understood so far.
After reviewing BV formalism and introducing Q- and QP-manifolds, 
relations between Q- and QP-manifolds and underling geometry are discussed. 
Based on these observations, BV action functionals $S_{BV}$ are constructed 
by the geometric quantities such as torsions and curvatures.

\section{Necessity of BV Formalism}\label{BVformalism}
Here, necessity of the BV formalism is explained in a basic example.

For a gauge field $A$, the action functional of the Yang-Mills theory on 
a four-dimensional manifold $\Sigma_4$ is 
\begin{eqnarray}
&& S = - \frac{1}{4} \int_{\Sigma_4} \mathrm{tr}(F \wedge * F),
\end{eqnarray}
where $F = \rd A + A \wedge A$ is a field strength.
Here $A = A_{\mu}^a \rd \sigma^{\mu} e_a$ is a connection $1$-form 
taking value on a Lie algebra $\mathfrak{g}$.
$\sigma^{\mu}$ is a local coordinate on $\Sigma_4$, and
$e_a$ is a basis of $\mathfrak{g}$.
Applying the variational principle to the action functional $S$, Euler-Lagrange equations which give equations of motion of physics are calculated.

The action functional $S$ has gauge invariance under 
the following infinitesimal gauge transformation:
\begin{eqnarray}
&& \delta_{\epsilon} A = \rd \epsilon + [A, \epsilon],
\label{gaugetransf}
\end{eqnarray}
where $\epsilon \in C^{\infty}(\Sigma_4, \mathfrak{g})$ is a gauge parameter,
which is a function on $\Sigma_4$ taking value on a Lie algebra $\mathfrak{g}$.
The gauge invariance means that that action functional is invariant 
under the gauge transformation \eqref{gaugetransf} $\delta_{\epsilon} S = 0$
for every $\epsilon$.
A set of gauge transformations is a closed Lie algebra 
under the commutators of transformations,
\begin{eqnarray}
[\delta_{\epsilon}, \delta_{\epsilon^{\prime}}] A
= \delta_{[\epsilon, \epsilon^{\prime}]} A,
\qquad \mbox{(off-shell)}
\nonumber
\end{eqnarray}
without using equations of motion,
where $[\delta_{\epsilon}, \delta_{\epsilon^{\prime}}] = 
\delta_{\epsilon} \delta_{\epsilon^{\prime}} - \delta_{\epsilon^{\prime}} \delta_{\epsilon}$.

The BRST quantization is a standard quantization procedure of the Yang-Mills theory. There are a lot of historical developments like gauge fixings, Gupla-Bleuler, Faddev-Popov, Becchi-Rouet-Stora-Tyutin (BRST), Kugo-Ojima, etc.
One can refer to some textbooks of a quantum field theory, for instance, \cite{Peskin:1995ev}.

We construct an operator of which square is zero and induced from the gauge transformation and make a consistent quantization by considering a cohomology called the BRST cohomology. For it, change a gauge parameter $\epsilon(\sigma)$ to a Grassmann odd field called the Faddeev-Popov (FP) ghost $c(\sigma)$. The corresponding transformation is denoted by $s$,
\begin{eqnarray}
&& 
s A = \rd c + [A, c].
\nonumber
\end{eqnarray}
The action functional is invariant under the BRST transformation $s S = 0$.
Ghost numbers are denoted by $\gh$ and defined as 
$\gh \, A = 0$ and $\gh \, c = 1$.
Then, $s$ is of ghost number one.
If we define $s c = -\frac{1}{2} [c, c]$, then we obtain
$s^2 A = s^2 c = 0$ without using equations of motion.

In order to make the gauge fixing,
we introduce the Grassmann odd antighosts $\bar{c}$ 
and the Grassmann evern Nakanishi-Lautrap field $b$ satisfying
the BRST transformations, $s \bar{c} = i b$ and $s b = 0$.
The gauge fixed action functional is 
given by adding gauge fixing terms to the action functional $S$
under the requirement $s S_{q}=0$.
It is 
\begin{eqnarray}
S_{q}
&=& \int_{\Sigma} \left(\mathrm{tr}(F \wedge * F)
+ b * \rd * A + * \frac{\alpha}{2} \mathrm{tr}(b b)
- i \bar{c} * \rd * D c \right).
\label{gaugeYM}
\end{eqnarray}
Here $\alpha$ is a constant, and $*$ is the Hodge star.
For consistency of the quantization after the gauge fixing,
$S_q$ must be invariant under the BRST transformation, $s S_{q}=0$.
It is concretely checked in Eq.~\eqref{gaugeYM}.
The partition function,
\begin{eqnarray}
Z &=& \int \calD A \calD b \calD c \calD \bar{c} e^{\frac{i}{\hbar} S_q},
\end{eqnarray}
is consistent with BRST (gauge) transformations.
Physical states are defined by the cohomology class
of the complex of all fields and ghosts, where the differential $Q_{BRST}$
is one induced from the BRST transformation $s$.
It means that physical states $|\mbox{phys}\rangle$ are cohomology classes
satisfying $Q_{BRST}|\mbox{phys}\rangle=0$,
where $Q_{BRST}$ a Grassmann odd charge operator for BRST transformations
\cite{Kugo:1979gm}.
Note that $Q_{BRST}$ is Grassmann odd and $Q_{BRST}^2=0$.
Since $Q_{BRST}^2=0$ without EOMs,
the condition $Q_{BRST}|\mbox{phys}\rangle=0$ is consistent.

In general constraint systems in the Hamiltonian mechanics, or equivalently, general gauge invariant Lagrangian mechanics, 
general consistency conditions of a gauge invariant classical theory are
\begin{eqnarray}
\delta_{\epsilon} S = 0 \ \ (\mbox{\rm off-shell}),
\qquad 
[\delta_{\epsilon}, \delta_{\epsilon^{\prime}}] 
= \delta_{[\epsilon, \epsilon^{\prime}]}
+ (\mbox{\rm equations of motion}).
\end{eqnarray}
Here, ``off shell'' means without using EOMs.
Especially, the theory is still consistent 
if a gauge algebra is closed using EOMs.
In this theory, the BRST transformation satisfies
\begin{eqnarray}
s S = 0 \ \ (\mbox{\rm off-shell}),
\qquad 
s^2 = (\mbox{\rm equations of motion}).
\end{eqnarray}
The BRST quantization procedure does not work in this case.
If we take gauge fixing for the quantization, 
gauge fixing terms are added in the action functional by introducing 
ghosts, antighosts, etc.
The gauge fixed action functional $S_q$ changes the EOMs from the classical action functional $S$. Thus,
$s^2= (\mbox{\rm equations of motion})$ does not hold after gauge fixing.
In general, corresponding BRST operator $Q_{BRST}$ does not satisfy 
$Q_{BRST}^2=0$, and 
the physical condition in the BRST formalism 
$Q_{BRST}|\mbox{phys}\rangle=0$ is inconsistent!

BV formalism makes the formulation consistent, and it is
the procedure to obtain an operator $Q$ such that $Q^2=0$ without EOMs.

\section{Batalin-Vilkovisky (BV) Formalism}
In this section, we explain BV formalism \cite{BV1, BV2}, especially, construction of 
a BV action functional $S_{BV}$ from a classical action functional $S$
satisfying
\begin{eqnarray}
s S_{BV} = 0, \ \ (\mbox{\rm off-shell}),
\qquad 
s^2 = 0, \ \ (\mbox{\rm off-shell}),
\nonumber
\end{eqnarray}
without changing physics.

For it, more auxiliary fields called \textit{antifields} are introduced.

\subsection{Poisson Sigma Model}
We explain the procedure of the BV formalism by an example, 
the Poisson sigma model (PSM) \cite{Ikeda:1993fh, Schaller:1994es}, 
which needs BV formalism for the quantization.

The Poisson sigma model is a sigma model from $\Sigma$ in two dimensions to a target manifold $M$ in $d$ dimensions.
Let $\sigma^{\mu}$ be a local coordinate on $\Sigma$
and $x^i$ be a local coordinate on $M$.
$i= 1, \ldots, d$ are target space indices.
Precisely, we consider the space of vector bundle morphisms 
$\Hom(T\Sigma, T^*M)$ between $T\Sigma$ and $T^*M$.

Let $\xzero:\Sigma \rightarrow M$ be a smooth map from $\Sigma$ to $M$
and $A = A_{\mu i}(\sigma) \rd \sigma^{\mu} \in \Omega^1(\Sigma, X^* T^*M)$ 
be a $1$-form taking a value on the pullback of $T^*M$.
The action functional of the Poisson sigma model is 
\begin{eqnarray}
S &=& 
\int_{\Sigma} \left(\bracket{A}{\rd_{\Sigma} X} 
+ (\pi \circ \xzero) (A, A) \right)
\nonumber \\
&=& 
\int_{\Sigma} \left(A_i \wedge \rd_{\Sigma} X^i 
+ \tfrac{1}{2} \pi^{ij}(\xzero) A_{i} \wedge A_{j} \right),
\label{PSMaction}
\end{eqnarray}
where $\bracket{-}{-}$ is the pairing of $TM$ and $T^*M$,
$\pi \in \Gamma(\wedge^2 TM)$ is a bivector field on $M$
and $\rd_{\Sigma}$ is the de Rham differential on $\Sigma$.
$\pi^{ij}(X)= - \pi^{ji}(X)$ is an antisymmetric tensor.

This model includes important examples in physics as
two-dimensional dilaton gravity (a deformation of the JT gravity)
\cite{Ikeda:1993aj, Grumiller:2002nm},
topological string theories, A-model, B-model \cite{Witten:1988xj, Witten:1991zz},
BF-type topological field theories \cite{Birmingham:1991ty}, etc.

Moreover, one of mathematical striking results is that 
the tree (disc) open string amplitude gives 
the Kontsevich's formula of the deformation quantization
on a Poisson manifold \cite{Cattaneo:1999fm, Kontsevich:1997vb}.

If $\pi^{ij}(\xzero)=0$, it is the Abelian BF theory.
If $\pi^{ij}$ is a linear function, $\pi^{ij}(\xzero)= f^{ij}{}_k \xzero^k$ witha constant $f^{ij}{}_k$, it is a non-Abelian BF theory 
with a Lie group gauge symmetry.

Let $\epsilon(\sigma) \in C^{\infty}(\Sigma, X^*TM)$ be a gauge parameter function taking a value in $X^*TM$.
$\partial_k = \tfrac{\partial}{\partial x^i}$ is a partial derivative on $M$.
The action functional is invariant under the following gauge transformations:
\begin{eqnarray}
&& 
\delta_{\epsilon} \xzero^i = - \pi^{ij}(\xzero) \epsilon_j,
\qquad 
\delta_{\epsilon} A_i = \rd_{\Sigma} \epsilon_i 
+ \partial_k \pi^{jk}(\xzero) A_j \epsilon_k,
\end{eqnarray}
if and only if a bivector field, i.e., the second-order antisymmetric tensor, $\pi$ satisfies the following identity:
\begin{eqnarray}
\frac{\partial \pi^{ij}}{\partial x^m} \pi^{mk}(x)
+ (ijk\ \mbox{cyclic})
=0,
\label{Jacobi}
\end{eqnarray}
The identity \eqref{Jacobi} is the Jacobi identity 
of the following Poisson bracket on the target space $M$,
\begin{eqnarray}
\{ f(x), g(x) \}_{PB} \equiv 
\frac{1}{2} \pi^{ij}(x) \frac{\partial f}{\partial x^i} 
\frac{\partial g}{\partial x^j},
\label{Poissonbracket}
\end{eqnarray}
for functions $f, g$ on $M$.
Therefore, if the target manifold $M$ is a Poisson manifold with the 
Poisson structure given by Eq.~\eqref{Poissonbracket},
the action functional is consistent.

There exists equivalent definition of the Poisson structure 
\eqref{Poissonbracket} based on a bivector field.
A multivector field is a section of exterior algebra of a tangent bundle 
$\Gamma(\wedge^{\bullet} TM)$.
An odd Poisson bracket $[-,-]_S$ called the \textit{Schouten bracket} is 
defined on $\Gamma(\wedge^{\bullet} TM)$ as a generalization of 
the Lie bracket of vector fields.
For a bivector field $\pi := \tfrac{1}{2} \pi^{ij}(x) \partial_i \wedge \partial_j$, $\pi^{ij}(x)$ gives a Poisson structure if and only if 
$[\pi, \pi]_S=0$.

The algebra of gauge transformations satisfies
\begin{eqnarray}
&& [\delta_{\epsilon_1}, \delta_{\epsilon_2}]
\xzero^i = \delta_{[\epsilon_1, \epsilon_2]} \xzero^i,
\\
&& [\delta_{\epsilon_1}, \delta_{\epsilon_2}]
A_{\mu i} = \delta_{[\epsilon_1, \epsilon_2]} A_{\mu i} 
+ \epsilon_{1j} \epsilon_{2k} 
\frac{\partial \pi^{jk}}{\partial \xzero^i \partial \xzero^l}(X)
\frac{\delta S}{\delta A_{\mu l}},
\label{gaugeA}
\end{eqnarray}
where $\tfrac{\delta S}{\delta A_{\mu l}}$ is the left-hand side of 
the Euler-Lagrange equation given from the variation of $S$
with respect to $A$.
Thus, it is zero if 
fields satisfy EOMs.
Thus, Eq.~\eqref{gaugeA} shows that the algebra of gauge transformations
is closed only using EOMs, not the BRST formalism but the BV formalism 
is needed to quantize the theory. Such an algebra is called an open algebra.

\subsection{BV Formalism of Poisson Sigma Model}\label{BVPSM}
The procedure to construct the BV action functional is as follows.

At first, replace the gauge parameter $\epsilon_i$ with the Grassmann 
odd FP ghost $c_i$ of ghost number one.
Original classical fields $X^i$ and $A_i$ are of ghost number zero.
Then,  gauge transformations $\delta_{\epsilon}$ are replaced with
BRST transformations $s$ of degree one.
Define the BRST transformation of the FP ghost $c_i$ as 
$s c_i= -\tfrac{1}{2} \tfrac{\partial \pi^{jk}}{\partial X^i} c_j c_k $.
Then, BRST transformations satisfy $s^2 X = 0$, $s^2 c = 0$, and
$s^2 A = 0$ using EOMs (on-shell). 

Next, introduce \textit{antifields} $\Phi^* = (X^*_i, A^{*i}, c^*_i)$ 
for all the classical fields and ghosts $\Phi = (X^i, A_i, c_i)$ 
in the BRST formalism.
Assign ghost numbers for each antifield such that 
$\gh \ \Phi + \gh \ \Phi^* = -1$.
If the ghost number is even, the field is Grassmann even,
and if the ghost number is odd, the field is Grassmann odd.
$(X^*_i, A^{*i}, c^*_i)$ are of ghost number $(-1, -1, -2)$.

Next, we introduce an odd Poisson bracket called \textit{BV brackets},
or the antibracket, as
\begin{align}
\sbv{F}{G} \equiv 
\sum_{\Phi} \int_{\Sigma} \left(F \tfrac{\rdif}{\partial \Phi(\sigma)}
\tfrac{\ld }{\partial \Phi^*(\sigma^{\prime})}  G
- 
F \tfrac{\rdif}{\partial \Phi^*(\sigma)}
\tfrac{\ld}{\partial \Phi(\sigma^{\prime})} G
\right)
\delta^2(\sigma-\sigma^{\prime})
\label{BVbracket}
\end{align}
for every functional $F, G$ of $\Phi$ and $\Phi^*$, 
where $\frac{\rdif}{\partial \Phi^*(\sigma)}$ is a left functional differential
and $\frac{\ld}{\partial \Phi(\sigma^{\prime})}$ is a right functional differential,
i.e., differential with proper sign factors since $\Phi$ and $\Phi^*$ are 
even and odd fields.
BV brackets for functions $f(\Phi, \Phi^*)$ of fields are defined by 
substituting the integration of mutliplying a test function 
$\tau(\sigma)$,
$F = \int_{\Sigma} \tau(\sigma) f(\Phi(\sigma), \Phi^*(\sigma)),
$
to Eq.~\eqref{BVbracket}.

The Batalin-Vilkovisky (BV) action functional $S_{BV}$ is an
extension of the classical action functional $S$ and 
has the following expansion:
\begin{eqnarray}
S_{BV} = S + (-1)^{\gh \Phi} \int_{\Sigma} \Phi^* s \Phi
+ S_2(\Phi^{*2}) + S_3(\Phi^{*3}) + \ldots.
\end{eqnarray}
Here the zeroth order and first order of $\Phi^*$ are fixed
by the classical action functional and BRST transformations.
Nontrivial terms are the second order and higher.
Higher terms are determined by imposing the condition,
\begin{align}
\sbv{S_{BV}}{S_{BV}}=0,
\end{align}
without using EOMs, which is called the \textit{classical master equation}.
%
The terms $S_2, S_3, \ldots$ are calculated order by order.
In the PSM, the solution $S_{BV}$ is 
\begin{align}
S_{BV} &= \int_{\Sigma} \left[A_{i} \wedge \rd_{\Sigma} \xzero^i 
+ \tfrac{1}{2} \pi^{ij}(\xzero) A_{i} \wedge A_{j} 
\right.
\nonumber \\ 
&
\left.
- X^{+ i} \pi^{ij} c_j
+ A^{+ i} \wedge \left(\rd c_{i} + \tfrac{\partial \pi^{jk}}{\partial X^i} A_{j} c_k \right)
+ \tfrac{1}{2} \tfrac{\partial \pi^{jk}}{\partial X^i} c^{+ i} c_j c_k
\right.
\nonumber \\ 
& 
\left.
- \tfrac{1}{4} \tfrac{\partial^2 \pi^{kl}}{\partial X^i \partial X^j}
A^{+i} \wedge A^{+j} c_k c_l
\right],
\label{BVfunctional}
\end{align}
where 
$\Phi^+$ is the Hodge dual of $\Phi^*$, i.e., 
$A_i \equiv \rd \sigma^{\mu} A_{\mu i}$, 
$A^{+i} \equiv \rd \sigma^{\mu} \epsilon_{\mu\nu} A^{*\nu i}$,
$\xzero^{+}_i \equiv \frac{1}{2} \rd \sigma^{\mu} \wedge \rd \sigma^{\nu} 
\epsilon_{\mu\nu} \xzero^{*}_{i}$, 
$c^{+i} \equiv \frac{1}{2} \rd \sigma^{\mu} \wedge \rd \sigma^{\nu} 
\epsilon_{\mu\nu} c^{*i}$.
$\epsilon_{\mu\nu}$ is the completely antisymmetric tensor 
on $\Sigma$ such that $\epsilon_{01}=1$. 


In the BV formalism, the BRST transformation is defined by
\begin{eqnarray}
s F[\Phi, \Phi^*] \equiv \sbv{S_{BV}}{F[\Phi, \Phi^*]},
\end{eqnarray}
for every functional $F$.
$S_{BV}$ is gauge invariant
because $s S_{BV}=\sbv{S_{BV}}{S_{BV}}$$=0$.
Moreover, $s^2=0$ 
\begin{eqnarray}
s^2 F &=& \sbv{S_{BV}}{\sbv{S_{BV}}{F}} 
= \frac{1}{2} \sbv{\sbv{S_{BV}}{S_{BV}}}{F} = 0.
\end{eqnarray}
for every functional $F$.

\subsection{Quantum BV Formalism}
Though we do not discuss the quantum BV formalism in this article,
it is briefly explained in this section.

In the path integral quantization,
we calculate the partition function,
\begin{eqnarray}
Z = \int_{\calL} \calD \Phi \ e^{\frac{i}{\hbar} S_{BVq}},
\end{eqnarray}
where 
$\calL$ is a Lagrangian submanifold with respect to the graded symplectic structure induced from the BV bracket \eqref{BVbracket}
of the space of fields $(\Phi, \Phi^*)$.
$S_{BVq}$ is the quantum action functional adding antighost $\bar{c}$, the NL field $b$, and their antifields.
Choice of $\calL$ corresponds to gauge fixing.
The partition function $Z$ must be invariant under infinitesimal changing of 
the Lagrangian submanifold, $\calL^{\prime} = \calL + \delta \calL$,
which means that the partition function $Z$ must not depend on choice of gauge.
The condition induces the following condition for $S_{BV q}$
called the quantum master equation,
\begin{eqnarray}
- 2 i \hbar \Delta S_{BV q} + \sbv{S_{BV q}}{S_{BV q}} =0.
\end{eqnarray}
Here $\Delta$ is the \textit{BV Laplacian} defined by $\Delta \equiv \sum_{\Phi} \int_{\Sigma} \tfrac{\ld}{\partial \Phi} \tfrac{\ld}{\partial \Phi^*}$,
which satisfies $\Delta^2=0$.

\section{Q-Manifolds and QP-Manifolds}\label{QP}
Mathematical structures in BV formalism 
in the physical theory in the previous section \ref{BVPSM}
are summarized as a Q-manifold or a QP-manifold.

BV formalism consists of three structures.
Fields, ghosts, and antifields are formulated as a \textit{graded manifold}.
A BV bracket is an odd Poisson bracket induced from 
a \textit{graded symplectic form}.
A BV action functional and the classical master equation 
are \textit{homological vector field} and its Hamiltonian function.
Three structures are called a QP-manifold, or a differential graded (dg)
symplectic manifold.

\begin{defn}
A \textit{graded manifold} $\calM = (M, \calO_M)$ on a smooth manifold $M$ 
is a ringed space which structure sheaf 
$\calO_M$ is $\BZ$--graded commutative algebras over
$M$,
locally isomorphic to $C^{\infty}(U) \otimes S^{\bullet}(V)$,
where $U$ is a local chart on $M$, 
$V$ is a graded vector space, and $S^{\bullet}(V)$ is a free 
graded commutative ring on $V$. 
\end{defn}
Grading is called \textit{degree}.
We denote the structure sheaf by $\calO_M = C^{\infty}(\calM)$
and degree of a homogeneous element $f$ by $|f|$.
If degrees are nonnegative, a graded manifold is called a \textit{N-manifold}.

The product of two homogeneous elements $q$ and $p$ of a graded manifold
satisfies $qp = (-)^{|q||p|} pq$.

A derivation on $C^{\infty}(\calM)$, i.e., a map on $C^{\infty}(\calM)$
satisfying the Leibniz rule is called a graded vector field.
A graded vector field $Q$ is called a 
\textit{homological vector field} if it is of degree $+1$ and $Q^2=0$.

\begin{defn}
A graded manifold with a homological vector field $(\calM, Q)$
is called a \textit{Q-manifold}, or differential graded (dg) manifold.
\end{defn}
A graded $1$-form is also defined as a map 
from the space of graded vector fields to $C^{\infty}(\calM)$,
and a graded differential form is defined by taking graded wedge products.

A graded symplectic form $\omega$ of degree $n$ on $\calM$ is a 
nondegenerate closed $2$-form of degree $n$.
A graded symplectic form induces a graded Poisson bracket $\sbv{-}{-}$ of degree $-n$ like the formula in normal symplectic geometry.
%
\begin{defn}
A triple $(\calM, \omega, Q)$ is called 
a \textit{QP-manifold}, or a dg symplectic manifold of degree $n$
if $\calL_Q \omega =0$.
Here $(\calM, Q)$ is a Q-manifold, and $\omega$ is a graded symplectic form.
\end{defn}

If degree $n \neq 0$, there exists a homological function
$\Theta \in C^{\infty}(\calM)$ of degree $n+1$ such that
$
Q(-) = \sbv{\Theta}{-}
$
\cite{Roytenberg:2006qz}.
$\Theta$ is a Hamiltonian function for a vector field $Q$
with respect to the symplectic form $\omega$.
$Q^2=0$ is equal to the equation,
$
\sbv{\Theta}{\Theta} =0
$.

The integration on the graded manifold is called the \textit{Berezin integral}.
Let $T[1]\BR^n \simeq \BR^n \times \BR^n[1]$ be the graded Euclidean space,
where $[1]$ means that degree of the space is shifted by one.
Take local coordinates $(\sigma^{\mu}, \theta^{\mu})$ 
on $\BR^n \times \BR^n[1]$,
where $\sigma^{\mu}$ is a coordinate of degree zero and 
$\theta^{\mu}$ is a coordinate of degree one.
Then the Berezin integral is denoted by
\begin{eqnarray}
&& \int_{T[1]\BR^n} \rd^n \sigma \rd^n \theta f(\sigma, \theta).
\end{eqnarray}
$\rd \sigma$ is a normal integration.
$\rd \theta$ is the integration of odd degree variables and is defined by
\begin{eqnarray}
&& \int \rd \theta (f_1(\sigma) + f_2(\sigma) \theta)= f_2(\sigma).
\end{eqnarray}

In order to define the Berezin integration on a graded manifold,
the transformation formula of the integration measure 
under change of variables,
$(\sigma, \theta) \mapsto (\sigma', \theta')$,
is needed.
Instead of the Jacobian in the normal integration,
the following Berezin matrix,
$\frac{\partial (\sigma', \theta')}
{\partial (\sigma, \theta)}
= \begin{pmatrix}
A & B \\
C & D \\
\end{pmatrix}
$
and the Berezinian,
$\mathrm{Ber} = \det(A - BD^{-1}C) (\det D)^{-1}
$,
appears in change of the integration measure. \cite{Manin}

\subsection{QP-Manifolds in the PSM}\label{QPPSM}
The space of (classical) BV formalism is 
a QP-manifold of degree $-1$ on the mapping 
space of two graded manifolds.
Here a graded manifold $\calM$ is a space of fields, ghosts, and antifields
$(\Phi, \Phi^*)$,
a graded symplectic form of degree $-1$ gives a BV bracket, and
a Hamiltonian function $\Theta$ is the BV action functional $S_{BV}$.

The QP-manifold of the Poisson sigma model is as follows.
The space of BRST fields is 
\begin{eqnarray}
\cM_{BRST} := \{ \left(X \colon \Sigma \to M, \, A \in \Omega^1(\Sigma,X^*T^*M), \, c \in C^\infty(\Sigma, X^*T^*[1]M)\right)\}\, .
\end{eqnarray}
The graded manifold is a graded cotangent bundle,
\begin{equation} 
\cM_{BV} := T^*[-1]\cM_{BRST}. \, 
\label{MBV}
\end{equation}
The graded symplectic form is
\begin{eqnarray}
\omega = \int_{\Sigma} 
\left(\delta X^i \wedge \delta X^+_i + \delta A_i \wedge \delta A^{+i}
+ \delta c_i \wedge \delta c^{+i}\right).
\label{componentBVsymplecticform}
\end{eqnarray}
The homological function is the BV action functional $S_{BV}$
in Eq.~\eqref{BVfunctional}.

\subsection{AKSZ Sigma Models}\label{QPPSM}
In the BV formalism in the Poisson sigma model, 
the BV action functional is simplified by considering superfields.

The two-dimensional manifold $\Sigma$ is extended to 
the graded tangent bundle $T[1]\Sigma$.
Take local coordinates $(\sigma^{\mu}, \theta^{\mu})$ on $T[1]\Sigma$ of degree $(0,1)$, where $\sigma^{\mu}$ is a coordinate on $\Sigma$
and $\theta^{\mu}$ is one on $T[1]$.

All the fields are combined into two-dimensional superfields as follows:
\begin{eqnarray}
\bx^i (\sigma, \theta)
&:= & 
\xzero^i (\sigma) - \theta^{\mu} A_{\mu}^{+i} (\sigma)
+ \tfrac{1}{2} \theta^{\mu} \theta^{\nu} c_{\mu\nu}^{+i} (\sigma),
\label{2Dsuperfield1} \\
\ba_i (\sigma, \theta) 
&:=&
- c_i (\sigma) + \theta^{\mu} A_{\mu i} (\sigma)
+ \tfrac{1}{2} \theta^{\mu} \theta^{\nu} \xzero_{\mu\nu i}^{+} (\sigma).
\label{2Dsuperfield2}
\end{eqnarray}
The graded manifold is 
${\mathcal{M}}_{BV} \cong \underline{\mathrm{Hom}}(T[1] \Sigma,T^*[1]M)$,
which is equivalent to $\cM_{BV}$ in Eq.~\eqref{MBV}.
The graded symplectic form in Eq.~\eqref{componentBVsymplecticform} 
is written as
\begin{eqnarray} 
\omega &=& \int_{T[1]\Sigma} \rd^2 \sigma \rd^2 \theta \,
\delta \bx^i \delta \ba_i.
\end{eqnarray}
The BV action functional in Eq.~\eqref{BVfunctional} is equivalently
written as
\begin{eqnarray}
S_{BV} 
&=& 
\int_{T[1]\Sigma} 
\!\!\!\!
\rd^2 \sigma \rd^2 \theta \,
\left[\ba_i \, \sd \bx^i 
+ \tfrac{1}{2} \pi^{ij}(\bx)\, \ba_i \ba_{j} \right].
\label{superfieldBVaction}
\end{eqnarray}

\section{Geometry of Q-Manifolds and QP-Manifolds}
In this section, we explain some important examples 
of Q-manifolds and QP-manifolds.
Examples in this section are graded manifolds in finite dimensions, 
which are basic examples of QP-manifolds.
AKSZ sigma models are directly induced from such QP-manifolds 
in finite dimensions.

\subsection{Lie Algebras}
Let $\mathfrak{g}$ be a vector space.
We consider a graded vector space $\calM = T^*[2]\mathfrak{g}[1] 
\simeq \mathfrak{g}[1] \oplus \mathfrak{g}^*[1]$.
Take odd coordinates of 
$\mathfrak{g}[1]$ and $\mathfrak{g}^*[1]$, $c^a$ and $b_a$.
Choose the canonical symplectic form,
$\omega = \delta c^a \wedge \delta b_a$.
Then, the nontrivial Poisson bracket is $\sbv{c^a}{b_b} = \delta^a_b$.

The homological function $\Theta$ is of degree three.
We take $\Theta = \frac{1}{2} C_{ab}^c c^a c^b b_c$,
where $C_{ab}^c$ is a constant. 
A homological condition of $\Theta$, $\sbv{\Theta}{\Theta} = 0$,
is equal to the identity of the structure constant $C_{ab}^c$
given by the Jacobi identity of the Lie bracket defined by
$[b_a, b_b] := C_{ab}^c b_c$.
Therefore, the QP-structure induces a Lie algebra structure on $\mathfrak{g}$.

$C^{\infty}(T^*[2]\mathfrak{g}[1])
\simeq \wedge^{\bullet} (\mathfrak{g} \oplus \mathfrak{g}^*)$
with $Q$ is equivalent to the Chevalley-Eilenberg complex of 
the Lie algebra $\mathfrak{g}$.

\subsection{Q-Manifolds of Degree One and Lie Algebroids}\label{Q1LA}
Let $E$ be a vector bundle over a smooth manifold $M$.

We consider a shifted vector bundle $E[1]$.
Let $(x^i, q^a)$ be a local coordinate on $E[1]$ of degree $(0, 1)$.

Let $Q$ be a vector field of degree one on $E[1]$.
A general form of a degree one vector field is
\begin{eqnarray}
Q = \rho^i_a(x) q^a \frac{\partial}{\partial x^i}
- \frac{1}{2} C_{bc}^a(x) q^b q^c \frac{\partial}{\partial q^a},
\end{eqnarray}
where $\rho^i_a(x)$ and $ C_{bc}^a(x)$ are local functions.
The homological condition $Q^2=0$ induces identities
on $\rho^i_a(x)$ and $ C_{bc}^a(x)$.
These identities give a Lie algebroid structure on $E$.
\begin{prop}\label{Q1mfdLA}
\cite{Vaintrob}
A Q-manifold $(E[1], Q)$ induces a Lie algebroid structure on $E$.
\end{prop}
\begin{defn}
A \textit{Lie algebroid} $(E, \rho, [-,-])$ is a vector bundle 
$E$ over $M$ with
a bundle map $\rho: E \rightarrow TM$ called the anchor map, 
and a Lie bracket
$[-,-]: \Gamma(E) \times \Gamma(E) \rightarrow \Gamma(E)$
satisfying the Leibniz rule,
\begin{eqnarray}
[e_1, fe_2] &=& f [e_1, e_2] + \rho(e_1) f \cdot e_2,
\nonumber
\end{eqnarray}
{where $e_i \in \Gamma(E)$ and $f \in C^{\infty}(M)$.}
\end{defn}
%
If we define $\rho(e_a) := \rho^i_a(x) \partial_i$ and 
$[e_a, e_b] := C_{ab}^c(x) e_c$ for the basis $e_a$ of $E$,
Proposition \ref{Q1mfdLA} is proved.

\begin{ex}[Lie algebras]
Let a manifold $M$ be one point $M = \{pt \}$. 
Then it is a Lie algebra $\mathfrak{g}$.
\end{ex}

\begin{ex}[Tangent Lie algebroids]
$E=TM$ and $\rho = \mathrm{id}$, 
$[-,-]$ is a normal Lie bracket on the space of vector fields $\mathfrak{X}(M)$.
\end{ex}

\subsection{QP-Manifolds of Degree One and Poisson Structures}
A QP-manifold of degree one is equivalent to a Poisson structure.

Take the shifted cotangent bundle $\calM= T^*[1]M$ of a smooth manifold $M$
with local coordinates $(x^i, \xi_i)$ of degree $(0,1)$.
The graded manifold $T^*[1]M$ has the canonical graded symplectic form 
$\omega_{can} = \delta x^i \wedge \delta \xi_i$ of degree one.
Since a homological function is of degree two, a general form is
\begin{eqnarray}
\Theta 
= \frac{1}{2} \pi^{ij}(x) \xi_i \xi_j,
\end{eqnarray}
The homological condition $\sbv{\Theta}{\Theta}=0$ give Eq.~\eqref{Jacobi}, 
which gives the Poisson bracket \ref{Poissonbracket}.
Therefore, the following proposition is obtained.
\begin{prop}
A QP-manifold $(T^*[1]M, \omega, Q)$ induces a Poisson structure on $M$.
\end{prop}
The structure of the BV formalism of the Poisson sigma model 
in Section \ref{BVPSM} is directly connected to this proposition.

\subsection{QP-Manifolds of Degree Two and Courant Algebroids}\label{Q2CA}
Next, we consider a geometric structure induced from 
a QP-manifold of degree two.

Let $E$ be a vector bundle over a smooth manifold $M$.
A typical example of a graded manifold of degree two is 
$\calM = T^*[2]E[1]$,
where $E$ is a vector bundle on $M$.
In this case, the structure sheaf $\calO_M = C^{\infty}(\calM)$
is not described by a space of sections of a vector bundle.
Assume a fiber metric $\bracket{-}{-} = k(-,-)$ to identify $E$ and $E^*$.
A local coordinate is $(x^i, \qone^a)$ of degree $(0,1)$, and 
the conjugate coordinate is $(\xi_i, k_{ab} \qone^b)$
of degree $(2,1)$.
Here $x^i$ is a local coordinate on $M$, and 
$\qone^a$ is a local coordinate of the fiber of $E[1]$.

\begin{ex}[$bc$-$\beta\gamma$ system in superstring theory]
A QP-manifold $T^*[2]E[1]$ appearing in a physical theory is 
the $bc$-$\beta\gamma$ ghost system in the superstring theory.
\cite{Ekstrand:2011cq}
For the BRST quantization of the superstring sigma model 
in two-dimensional manifold, we need to introduce 
Grassmann odd ghosts for bosonic string theory $b^a$ and $c_a$
and Grassmann even ghosts for super string theory $\beta_i$ and $\gamma^i$.
$(b^a, c_a, \beta_i, \gamma^i)$ are of degree $(1, 1, 2, 0)$.
$(b^a, c_a)$ corresponds to $\eta^a$, $\beta_i$ to $\xi_i$,
and $\gamma^i$ to $x^i$.
Graded Poisson brackets of each field are
$\{b^a, c_b \}= \delta^a_b$, $\{\beta_i, \gamma^j \}= - \delta_i^j$.
induced from $\omega$.

The graded symplectic form $\omega$ must be invariant under
local coordinate transformations on a QP-manifold of degree $2$
induced from the definition of structure sheaf.
They are canonical transformations on the $bc$-$\beta\gamma$ system,
Local coordinate transformations are
\begin{eqnarray}
\gamma^{\prime i} &=& \gamma^{\prime i}(\gamma),
\qquad b^{\prime a} = M^a_b(\gamma) b^{b},
\qquad 
c^{\prime}_a = M^b_a(\gamma) c_{b},
\label{Q2transf1}
\\
\beta^\prime_i &=& \frac{\partial \gamma^j}{\partial \gamma^{\prime i}} \beta_j
+ \frac{1}{2} M^c_b \frac{\partial M^d_c}{\partial \gamma^{\prime i}} b^{b} c_{d},
\label{Q2transf2}
\end{eqnarray}
where $\gamma^{\prime i}(\gamma)$ is arbitrary function of $\gamma$,
and $M^a_b(\gamma)$ is a transition function of the fiber of $E$.
Eqs.~\eqref{Q2transf1} and \eqref{Q2transf2} are equations of 
local coordinate transformations of the structure sheaf in $T^*[2]E[1]$
\cite{Roy01}.
The transformation of $\beta_i$ is not realized by a fiber bundle over $M$.

\end{ex}

Since $T^*[2]E[1]$ is a cotangent bundle, the graded manifold has 
a canonical graded symplectic form of degree two,
\begin{eqnarray}
\omega = \inner{\delta x}{\delta \xi} + 
\bracket{\delta \qone}{\delta \qone}
= \delta x^i \wedge \delta \xi_i + 
\frac{1}{2} \delta \qone^a \wedge \delta (k_{ab} \qone^b),
\end{eqnarray}
where $\inner{-}{-}$ is the pairing between $TM$ and $T^*M$.
A general homological function of degree $3$ is
\begin{align}
\Theta &= 
\rho^i{}_a(x) \xi_i \qone^a + \frac{1}{3!} C_{abc}(x) 
\qone^a \qone^b \qone^c,
\label{Q2homological}
\end{align}
where $\rho^i{}_a(x)$ and $C_{abc}(x)$ are local functions on $M$.
If the homological condition $\sbv{\Theta}{\Theta}=0$ is imposed,
we obtain identities of $\rho^i{}_a(x)$ and $C_{abc}(x)$.

Identities induce the following proposition.
\begin{prop}\label{QP2courant}
\cite{Roy01}
A QP-manifold of degree $2$ induces
a Courant algebroid structure on $E$.
\end{prop}
The Courant algebroid \cite{Courant, LWX} is defined as follows.
\begin{defn}\label{courantdefinition}
A Courant algebroid is a vector bundle $E \rightarrow M$,
and it has a nondegenerate symmetric bilinear form
$\bracket{-}{-}$ on the bundle, 
a bilinear operation $(-\circ-)= \courant{-}{-}$ 
on $\Gamma (E)$,
and a bundle map called an anchor map,
$\rho: E \longrightarrow TM$, satisfying the following properties:
%
\begin{eqnarray}
&& 1, \quad e_1 \circ (e_2 \circ e_3) = (e_1 \circ e_2) \circ e_3 
+ e_2 \circ (e_1 \circ e_3), 
  \label{courantdef1}
\\
&& 2, \quad \rho(e_1 \circ e_2) = [\rho(e_1), \rho(e_2)], 
  \label{courantdef2}
\\
&& 3, \quad e_1 \circ f e_2 = f (e_1 \circ e_2)
+ (\rho(e_1)f)e_2, 
  \label{courantdef3}
 \\
&& 4, \quad e_1 \circ e_2 = \frac{1}{2} {\calD} \bracket{e_1}{e_2},
  \label{courantdef4}
\\ 
&& 5, \quad \rho(e_1) \bracket{e_2}{e_3}
= \bracket{e_1 \circ e_2}{e_3} + \bracket{e_2}{e_1 \circ e_3},
  \label{courantdef5}
\end{eqnarray}
where 
$e_1, e_2$, and $e_3$ are sections of $E$, $f$ is a function on
$M$ and 
${\calD}$ is a map from the space of functions on $M$ to $\Gamma (E)$, 
defined as 
$\bracket{{\calD}f}{e} = \rho(e) f$.
\end{defn}
A bilinear operation $(-\circ-)= \courant{-}{-}$ is called the Dorfman bracket.

Correspondence between a QP-manifold of degree two and 
a Courant algebroid in Proposition \ref{QP2courant} is based on 
the \textit{derived bracket construction}.

The structure sheaf $\calO_M = C^{\infty}(\calM)$
is decomposed by degree as \\
$C^{\infty}(\calM)  = \sum_{i \geq 0} C_i(\calM)$,
where $C_i(\calM)$ is the space of functions of degree $i$.
Take the subspace of functions of degrees zero and one,
$C_0(\calM) \oplus C_1(\calM)$. $C_0(\calM) \oplus C_1(\calM)$ 
is a closed algebra by the bracket $\sbv{-}{-}$ and 
the derived bracket $\sbv{\sbv{-}{\Theta}}{-}$. 
It is easily checked by degree counting.
$C_0(\calM)$ is equivalent to the space of 
functions on $M$, $C^{\infty}(M)$.
$C_1(\calM)$ is the space of sections on $E$, $\Gamma(E)$.
A function of degree one $\alpha_a(x) \qone^a \in C_1(\calM)$
is identified by a section of $E^*$, $e = \alpha_a(x) e^a \in \Gamma(E^*)$, 
but $E^*$ is identified to $E$ by the fiber metric $\bracket{-}{-}$.

Operations on the Courant algebroid $E$ are defined 
by Poisson brackets and derived brackets.
For $f,g \in C_0(\calM)$, $e, e_1, e_2 \in C_1(\calM)$,
Poisson brackets on $C_0 \times C_0$ and $C_1 \times C_0$ are zero.
Brackets on $C_1 \times C_1$ gives the inner product
$\bracket{e_1}{e_2} = \sbv{e_1}{e_2}$ for $e_1, e_2 \in \Gamma(E)$.
Derived brackets of $C_0 \times C_0$ are zero from degree counting,
$0= \sbv{\sbv{f}{\Theta}}{g}$.
Derived brackets $C_1 \times C_0 \rightarrow C_0$ give the anchor map,
$\rho(e)f = - \sbv{\sbv{e}{\Theta}}{f}$.
Derived brackets $C_1 \times C_1 \rightarrow C_1$ give
the Dorfman bracket $\courant{e_1}{e_2} = - \sbv{\sbv{e_1}{\Theta}}{e_2} 
$.
Moreover, ${\calD}f$ in Definition \ref{courantdefinition} is 
given by ${\calD}f = \sbv{\Theta}{f}$.

All the identities in Definition \ref{courantdefinition} are proved 
from one identity, the homological condition $\sbv{\Theta}{\Theta} = 0$.
Conversely, if we assume a Courant algebroid structure $E$,
a QP-manifold structure on $T^*[2]E[1]$ is constructed.
Especially, a homological function $\Theta$ is given by 
Eq.~\eqref{Q2homological}.
Here for the basis of $E$, $e_a$, local functions are given by 
$\rho(e_a) = \rho^i_a(x) \partial_i$
and $\bracket{e_a}{e_b} = C_{abc}(x) k^{cd} e_d$.

For a QP-manifold of general $n$, an algebroid structure on a vector bundle $E$
is induced by the derived brackets construction similar to a Courant algebroid case.
An algebroid corresponding to the QP manifold of degree $n$ is called 
a \textit{$L_n$ algebroid}. A $L_1$ algebroid is a Poisson-Lie algebroid.
A $L_2$ algebroid is a Courant algebroid. 
A Lie algebroid has correspondence to a Lie groupoid as a generalization of 
correspondence between a Lie algebra and a Lie group.
A corresponding global object to a $L_n$ algebroid is 
called a \textit{$L_n$ groupoid}.

\subsection{Courant Sigma Model}
If a QP-manifold is given, a sigma model is constructed.
We can construct a sigma model with a Courant algebroid structure
from a QP-manifold of degree $2$.
\cite{Ikeda:2002wh, Roytenberg:2006qz}

The BV action functional of the Courant sigma model is defined 
on the mapping space, 
$\Map(T[1]\Sigma, T^*[2]E[1])$, where $\Sigma$ 
is a three-dimensional smooth manifold.
The BV action functional of the Courant sigma model is
\begin{eqnarray}
S&=& \int_{T[1]\Sigma} d^{3}\sigma d^{3}\theta
\ \left(- \bb_{i} \bbd \bx^{i}
+ \frac{1}{2} k_{ab} \ba^{a} \bbd \ba^{b}
\right.
\nonumber \\ && 
\left.
+ \rho^{i}_{a}(\bx)
\bb_{i} \ba^{a} 
+ \frac{1}{3!}
H{}_{abc}(\bx)
\ba^{a} \ba^{b} \ba^{c}
\right).
\label{3DCSM}
\end{eqnarray}
Here $(\bx^{i}, \ba^a, \bb_i)$ are 
local coordinates on the mapping space, i.e., superfields
of degree $(0, 1, 2)$, respectively.
Precisely, they are $\bx: T[1]\Sigma \rightarrow M$, 
$\ba \in \Gamma(T[1]\Sigma, \bx^* E[1])$, 
$\bb \in \Gamma(T[1]\Sigma, \bx^* T^*[2]M)$.
$(\bx^{i}, \ba^a, \bb_i)$ correspond to $(x^i, \eta^a, \xi_i)$ in
a QP-manifold of degree two in Section \ref{Q2CA}.
One can refer to \cite{Ikeda:2012pv} for detailed constructions.

\section{Geometric Construction of BV Action Functionals}\label{GeomBV}
A BV action functional $S_{BV}$ is usually complicated
and calculations are not easy. Moreover, 
geometric meanings of each term in $S_{BV}$ are not well understood.
In this section, we discuss relations of the BV action functional $S_{BV}$
to geometries of Lie algebroids and higher algebroids.
We mainly consider a Lie algebroid case as an example.
Generalizations to the Courant algebroid case are commented in 
the last part of the section.
One can refer to \cite{Ikeda:2019czt, Chatzistavrakidis:2021nom, Chatzistavrakidis:2022hlu, Chatzistavrakidis:2023lwo, Ikeda:2021rir}.

Let $E$ be a Lie algebroid over a smooth manifold $M$.
As discussed in Section \ref{Q1LA}, there is always a corresponding Q-manifold.

There are two differentials in Lie algebroids.
The first differential is the de Rham differential on $M$,
$\rd: \Gamma(\wedge^l T^*M) \rightarrow  \Gamma(\wedge^{l+1} T^*M)$.
The second differential is a differential ${}^E \rd$ on the exterior algebra 
of $E^*$, $\wedge^m E^*$.
Sections on $\wedge^m E^*$ are called \textit{$E$-differential forms}, and 
the second differential
is defined as follows. 
\begin{defn}
For an $E$-differential form $\alpha \in \Gamma(\wedge^m E^*)$,
a \textit{Lie algebroid differential}, or an \textit{$E$-differential}, 
${}^E \rd: \Gamma(\wedge^m E^*)
\rightarrow \Gamma(\wedge^{m+1} E^*)$ is defined by
\begin{align}
{}^E \rd \alpha(e_1, \ldots, e_{m+1}) 
&= \sum_{i=1}^{m+1} (-1)^{i-1} \rhoa(e_i) \alpha(e_1, \ldots, 
\check{e_i}, \ldots, e_{m+1})
\nonumber \\ & 
\!\!\!\!\!\!\!\!\!\!\!\!\!\!\!\!\!\!\!\!\!\!\!\!\!\!\!\!\!\!\!\!
+ \sum_{1 \leq i < j \leq m+1} (-1)^{i+j} \alpha([e_i, e_j], e_1, \ldots, \check{e_i}, \ldots, \check{e_j}, \ldots, e_{m+1}),
\label{LAdifferential}
\end{align}
where $e_i \in \Gamma(E)$.
\end{defn}
The $E$-differential satisfies $({}^E \rd)^2=0$.
It is regarded as a generalization of the de Rham differential 
and the Chevalley-Eilenberg differential on a Lie algebra.

Corresponding to two differentials, two connections are introduced.
The first connection is a normal vector bundle connection.
\begin{defn}
A \textit{connection} on a vector bundle $E'$ 
is a $\BR$-linear map
$\nabla: \Gamma(E') \rightarrow {\Gamma(E' \otimes T^*M)}$ 
satisfying 
\begin{eqnarray}
\nabla_v (f e') = f \nabla_v e' + (v f) e',
\nonumber
\end{eqnarray}
for $v \in \Gamma(TM)$, $e' \in \Gamma(E')$ and $f \in C^{\infty}(M)$.
\end{defn}
The second ``connection'' is defined if $E$ is a Lie algebroid.
\begin{defn}
An \textit{$E$-connection} on a vector bundle $E'$ 
with respect to a Lie algebroid $E$ is a $\BR$-linear 
map
${}^E \nabla: \Gamma(E') \rightarrow {\Gamma(E' \otimes E^*)}$ 
satisfying 
\begin{eqnarray}
{}^E \nabla_e (f e') = f {}^E \nabla_e e' + (\rho(e) f) e',
\nonumber
\end{eqnarray}
for $e \in \Gamma(E)$, $e' \in \Gamma(E')$ and $f \in C^{\infty}(M)$.
\end{defn}
The ordinary connection is regarded as an 
$E$-connection for $E=TM$, $\nabla = {}^{TM} \nabla$.

For a given vector bundle connection 
$\nabla:\Gamma(E) \rightarrow {\Gamma(E \otimes T^*M)}$ on $E$,
an $E$-connection called the \textit{basic $E$-connection} on $TM$, ${}^E \nabla: \Gamma(TM) \rightarrow {\Gamma(TM \otimes E^*)}$ is defined by
\begin{eqnarray}
{}^E \nabla_{e} v := \calL_{\rho(e)} v + \rho(\nabla_v e)
= [\rho(e), v] + \rho(\nabla_v e).
\nonumber
\nonumber
\end{eqnarray}
The \textit{basic $E$-connection} on $E$, ${}^E \nabla: \Gamma(E) \rightarrow {\Gamma(E \otimes E^*)}$ is defined by
\begin{eqnarray}
{}^E \nabla_{e} e' &:=& \nabla_{\rho(e')} e + [e, e'],
\nonumber
\nonumber
\end{eqnarray}
for $e, e' \in \Gamma(E)$.

From now on, we always take the basic $E$-connection 
as the $E$-connection ${}^E \nabla$.

Corresponding to two connections, a vector bundle connection $\nabla$
and an $E$-connection ${}^E \nabla$, 
two curvatures and two torsions are defined.
A \textit{curvature}, $R \in \Omega^2(M, E \otimes E^*)$,
and an \textit{$E$-curvature}, 
${}^E R \in \Gamma(\wedge^2 E^* \otimes E \otimes E^*)$ are defined by
\beqa
R(v, v^{\prime}) &:=& [\nabla_v, \nabla_{v^{\prime}}] - \nabla_{[v, v^{\prime}]}, 
\label{curv} \\
{}^E R(e, e^{\prime}) &:=& [\anabla_e, \anabla_{e^{\prime}}] - \anabla_{[e, e^{\prime}]}, 
\label{Ecurv}
\eeqa
where $v, v^{\prime} \in \mathfrak{X}(M)$ and $e, e^{\prime} \in \Gamma(E)$.

Though a torsion and an \textit{$E$-torsion} are defined,
we use only the \textit{$E$-torsion}, 
$T \in \Gamma(E \otimes \wedge^2 E^*)$, 
\begin{eqnarray}
&& T(e,e') =  \nabla_{\rho(e)} e' - \nabla_{\rho(e')} e - [e,e']
\end{eqnarray}
Moreover, very important one more curvature-like geometric 
quantity called a \textit{basic curvature} is defined,
which contains two connections.
The basic curvature,
$\baS \in \Omega^1(M, \wedge^2 E^* \otimes E)$
\cite{Blaom}, is defined by
\beqa
\baS(e, e^{\prime}) &:=& 
[e, \nabla e^{\prime}] - [e^{\prime}, \nabla e] - \nabla[e, e^{\prime}] 
- \nabla_{\rhoa(\nabla e)} e^{\prime} + \nabla_{\rhoa(\nabla e^{\prime})} e
\nonumber \\
&=& (\nabla T + 2 \mathrm{Alt} \, \iota_\rho R)(e, e^{\prime}),
\label{bcurv}
\eeqa
for $v, v^{\prime} \in \mathfrak{X}(M)$ and $e, e^{\prime} \in \Gamma(E)$.

Using geometric quantities, the BV action functional of 
the Poisson sigma model \eqref{BVfunctional} is given 
in the following geometric form for $E=T^*M$,
\begin{eqnarray} 
&&  S_{BV} = S_{BV}^\nabla
:=  \int_{\Sigma} \left[\langle A ,
\bbd \xzero \rangle + \tfrac{1}{2} (\pi \circ X)(A, A) 
\right.
\nonumber \\ && 
\left.
+ \langle A^{+} , \nabla c 
-  (T\circ X) (A, c) \rangle - (\pi\circ X) (\xzero^{+\!\nabla},  c) 
\right.
\nonumber \\ && 
\left.
- \tfrac{1}{2}\langle c^+, (T\circ X) (c, c) \rangle
+ 
\tfrac{1}{4} 
\langle A^+, (\baS \circ X)(A^+, c,c) \rangle
\right]. 
\label{GeomBVPSM}
\end{eqnarray}
Here $\xzero^{+\!\nabla}_i$ is the covariantized antifield of $X^i$
defined by
$
\xzero^{+\!\nabla}_i := \xzero^{+}_i + \Gamma_{ji}^k (A^{+j} \wedge A_k + c^{+ j} c_k)
$,
where $\Gamma_{ji}^k$ is an affine connection on $TM$.
Note that the BV action functional does not depend on 
the connection $\nabla$ (and ${}^E \nabla$), i.e., $S_{BV} = S_{BV}^{\nabla}$.
However, each term does.
$\sbv{S_{BV}}{S_{BV}} = 0$ is proven using fundamental identities of 
the Lie algebroid and the ``Bianchi'' identity of the basic curvature,
${}^E \nabla \baS=0$.

\section{Deformations of Action Functionals}
The Poisson sigma model is deformed as a consistent classical field theory.
Then, the BV action functional $S_{BV}$ is not a simple deformation of the 
BV action functional of the PSM. 
Thus, to determine the BV action functional is a nontrivial problem.

Let $\Xi$ be a three-dimensional manifold with two-dimensional 
boundary $\Sigma = \partial \Xi$.
Let $H$ be a closed $3$-form on $M$.
The classical action functional of the PSM \eqref{PSMaction}
is deformed by adding the $H$ term,
\begin{eqnarray}
S &=& 
\int_{\Sigma} \left(\bracket{A}{\rd_{\Sigma} X} 
+ (\pi \circ \xzero) (A, A) \right)
+ \int_{\Xi} X^*H.
\label{TPSMaction}
\end{eqnarray}
This theory is consistent as a classical field theory and 
invariant under gauge transformations, if the target manifold $M$ is
the twisted Poisson manifold \cite{Klimcik:2001vg, Severa:2001qm}.
\begin{defn}
Let $\pi \in \Gamma(\wedge^2 TM)$ and $H \in \Omega^3(M)$ be a closed $3$-form.
If $(\pi, H)$ satisfies
\begin{eqnarray}
&& \tfrac{1}{2}[\pi, \pi]_S 
= \bracket{\otimes^{3} \pi}{H},
\label{tPoisson1}
\nonumber
\end{eqnarray}
$\bracket{\otimes^{3} \pi}{H}$ is defined by 
$\bracket{\otimes^{3} \pi}{H}(\alpha_1, \alpha_2, \alpha_3)
:= H(\pi^{\sharp} (\alpha_1), \pi^{\sharp} (\alpha_2), 
\pi^{\sharp} (\alpha_3))$,
for $\alpha_i \in \Omega^1(M)$.
A bundle map $\pi^{\sharp}: T^*M \rightarrow TM$ is defined by
$\pi^{\sharp}(\alpha) = \pi(\alpha, -)$ for 
a $1$-form $\alpha \in \Omega^1(M)$.
\end{defn}

If $M$ has a twisted Poisson structure, $T^*M$ is a Lie algebroid.
The anchor map is defined by $\rho= -\pi^{\sharp}: T^*M \rightarrow TM$,
and the Lie bracket is defined by
\begin{eqnarray}
[\alpha, \beta]_{\pi,H} := \calL_{\pi^{\sharp} (\alpha)}\beta - \calL_{\pi^{\sharp} (\beta)} \alpha - \rd(\pi(\alpha, \beta))
+ \iota_{\pi^{\sharp}(\alpha)} \iota_{\pi^{\sharp}(\beta)} H,
\end{eqnarray}
for $\alpha, \beta \in \Omega^1(M)$.
Then, $(T^*M, [-, -]_{\pi, H}, -\pi^{\sharp})$ is a Lie algebroid.

We consider the BV action functional $S_{BV}$ of the classical 
action functional \eqref{TPSMaction} of the twisted PSM.
$S_{BV}$ of the classical theory \eqref{TPSMaction}
is not a simple deformation of the BV action functional \eqref{BVfunctional}
in the PSM.
However, the geometric BV formalism gives the BV action functional of the twisted theory by a simple extension.
This is an advantage of the geometric BV formalism.
$S_{BV}$ in the twisted theory is formally the same as 
the geometric BV action functional \eqref{GeomBVPSM}.
\begin{thm}\cite{Ikeda:2019czt}
The BV action functional of the twisted PSM is 
\begin{eqnarray} 
&&  S_{BV} = S_{BV}^\nabla
:=  \int_{\Sigma} \left[\langle A ,
\bbd \xzero \rangle + \tfrac{1}{2} (\pi \circ X)(A, A) 
\right]
+ \int_{T[1]\Xi} X^*H
\nonumber \\ && 
+ 
\int_{\Sigma} \left[
\langle A^{+} , \nabla c 
-  (T\circ X) (A, c) \rangle - (\pi\circ X) (\xzero^{+\!\nabla},  c) 
\right.
\nonumber \\ && 
\left.
- \tfrac{1}{2}\langle c^+, (T\circ X) (c, c) \rangle
+ 
\tfrac{1}{4} 
\langle A^+, (\baS \circ X)(A^+, c,c) \rangle
\right]
\label{GeomBVTPSM}
\end{eqnarray}
Formally we only add the term, $\int_{T[1]\Xi} X^*H$ in \eqref{GeomBVPSM}.
However, note that geometric quantities $\nabla$, $T$, $\baS$ in $S_{BV}$ are deformed from the Lie algebroid induced from the Poisson structure to one induced from the twisted Poisson structure.
\end{thm}

\section{BV of Twisted Courant Sigma Models}
AKSZ sigma models on $n+1$-dimensional manifold $\Sigma$
can be deformed by adding the term $\int_{\Xi} X^*H$ 
in the action functional. 
Here $H$ is a closed $n+2$-form $H$ on $M$
and $\Xi$ is an $n+2$-dimensional manifold with $n+1$-dimensional 
boundary $\Sigma = \partial \Xi$.
The theory is still consistent as a classical field 
theory with gauge symmetries.
However, the BV action functional is not one adding 
the $\int_{\Xi} X^*H$ term, nor simple deformations.

Recently, $n=2$ case has been analyzed in details \cite{Chatzistavrakidis:2022hlu, Chatzistavrakidis:2023otk}.

$\Xi$ is four dimensions, and $\Sigma$ is three dimensions.
The Courant sigma model \eqref{3DCSM} is deformed
by introducing a closed $4$-form $H$ on $M$.
We call it the \textit{twisted Courant sigma model}.
The classical action functional is
\begin{eqnarray}
S&=& \int_{\Sigma} d^{3}\sigma
\ \left(- B_{i} \bbd X^{i}
+ \frac{1}{2} k_{ab} A^{a} \bbd A^{b}
\right.
\nonumber \\ && 
\left.
+ \rho^{i}_{a}(X)
B_{i} A^{a} 
+ \frac{1}{3!}
H{}_{abc}(X)
A^{a} A^{b} A^{c}
\right)
+ \int_{\Xi} X^* H,
\label{3DTCSM}
\end{eqnarray}
where $X$ is a map from $\Xi$ to $M$,
$A \in \Omega^1(\Sigma, X^* E)$ is a $1$-form taking a value on the pullback of $E$, and $B \in \Omega^2(\Sigma, X^* T^*M)$ is a $2$-form taking a value on the pullback of $T^*M$.
If the vector bundle $E$ is a $4$-form twisted Courant algebroid, the theory is consistent as a classical field theory \cite{Hansen:2009zd}.

A $4$-form  twisted Courant algebroid is defined as follows.
\begin{defn} \cite{Vaisman:2004msa}
A \textit{pre-Courant algebroid} is a vector bundle $E$
with a pseudo-Euclidean inner product $\langle\cdot,\cdot\rangle$
and the anchor map $\rho:E\to TM$ 
together with a binary operation $\circ$ on $\Gamma(E)$ 
that satisfies the following axioms
for all $e, e',e'' \in \Gamma(E)$:
Here $\rho^{\ast}:T^{\ast}M\to E^{\ast}\simeq E$ is the transpose map.
\begin{eqnarray}
\rho\cdot\rho^{\ast}&=&0\,,
\qquad
\rho(e\circ e')=[\rho(e),\rho(e')]\,,
\nonumber 
\\
\langle e\circ e,e'\rangle&=&\tfrac 12 \rho(e')\langle e,e\rangle\,,
\qquad
\rho(e)\langle e', e'' \rangle= \langle e\circ e', e'' \rangle +\langle e', e\circ e''\rangle\,.
\nonumber
\end{eqnarray}
\end{defn} 
A \textit{4-form twisted Courant algebroid} is a pre-Courant algebroid with 
the Jacobiator given by
\bea 
e\circ(e'\circ e'')-(e\circ e')\circ e'' -e'\circ(e \circ e'')
= \rho^{\ast}H(\rho(e),\rho(e'),\rho(e''))\,.
\eea
Similar to Section \ref{GeomBV}, for a pre-Courant algebroid,
two connections, two torsions, two curvatures,
and the basic curvature are defined
\cite{Gualtieri:2007bq, Jotz, Chatzistavrakidis:2023otk}.
Especially, the $E$-torsion and the basic curvature are needed 
for the BV action functional $S_{BV}$.
Let $e, e' \in \Gamma(E)$, $v \in \mathfrak{X}(M)$.
The \textit{$E$-torsion},  $T \in \Gamma(\wedge^3 E^*)$,
and the \textit{basic curvature}, 
${}^E S \in \Omega^1(\wedge^2 E^* \otimes E^*)$, are defined by
\begin{eqnarray}
T(e_1,e_2,e_3)&=&\langle {}^{E}\nabla_{e_1}e_2- {}^E\nabla_{e_2}e_1 - [e_1,e_2]_{E}, e_3\rangle 
\nonumber \\ &&
+ \tfrac{1}{2} \left(\langle  {}^E \nabla_{e_3}e_1,e_2\rangle 
-\langle {}^E \nabla_{e_3}e_2,e_1\rangle \right),
\nonumber 
\\
{}^E S(e_1,e_2,e_3)v &=&
\langle \nabla_{v}[e_1,e_2]_{E}- [\nabla_v e_1,e_2]_{E}-[e_1,\nabla_v e_2]_{E}
\nonumber\\
&& -\nabla_{{}^E \nabla_{e_2}v}e_1+\nabla_{{}^E \nabla_{e_1}v}e_2 , e_3\rangle
 \,\,\nonumber\\ &&
+\tfrac{1}{2} \left(\langle\nabla_{{}^E \nabla_{e_3}v}e_1,e_2\rangle
-\langle\nabla_{{}^E \nabla_{e_3}v}e_2,e_1\rangle \right)\,,
\nonumber
\end{eqnarray}
in terms of the skew-symmetric Courant bracket $[e_1,e_2]_{E}
= \tfrac{1}{2} ([e_1,e_2]_{D} - [e_2,e_1]_{D})$.

The BV action functional is $\tS =S_0+S_1+S_2+S_3$ \cite{Chatzistavrakidis:2023lwo}, where
\bea 
S_0&=& \int \bigl\{-( B^{\scriptscriptstyle\nabla},F) +  \langle A,{\rm D}A\rangle-\tfrac{1}{3} {T}(A,A,A)+\int_{\Xi} X^{\ast}H \bigr\} \,,
\nonumber
\eea
\bea 
S_1&=&\int \bigl\{(X^{+\scriptscriptstyle\nabla}, \r(\e)) -\langle A^+, \DD\e+(\rho^\ast, \psi^{\scriptscriptstyle\nabla})-2{T}^\ast(-, A,\e) \rangle \nonumber\\
&+&\left(B^+, \DD\psi^{\scriptscriptstyle\nabla}+(\nabla\rho(A), \psi^{\scriptscriptstyle\nabla})+(\nabla\rho(\e), B^{\scriptscriptstyle\nabla})+{}^E S(A,A,\e)\right.\nonumber\\ 
&+& \left.\tfrac{1}{2}( H(\r(\e),\r(A))-R(\e,A))(-,F)+\tfrac{1}{6} H(-,\r(\e),F,F)\right)\nonumber\\ 
&+& 2\langle {\e}^+, (\rho, \widetilde{\psi}^{\scriptscriptstyle\nabla}) + {T}(-,\e, \e) \rangle 
+ (\widetilde{\psi}^{+}, (\nabla \rho(\e), \tilde{\psi}) + \tfrac{1}{3} {}^E S(\e, \e, \e)) \nonumber\\
&+&\left( \psi^+, {\rm D}\widetilde{\psi}^{\scriptscriptstyle\nabla}+(\nabla\r(A), \widetilde{\psi}^{\scriptscriptstyle\nabla})-(\nabla\r(\e), {\psi}^{\scriptscriptstyle\nabla})+{}^E S(A,\e,\e) \right.\nonumber\\
&+& \left. \tfrac{1}{4} (H(\r(\e),\r(e))-R(\e,\e))(-,F))\right) \bigr\}\,,
\nonumber 
\\
S_2&=& \int \bigl\{ -(B^+, - (\nabla \rho(A^{+*}), \widetilde{\psi}^{\scriptscriptstyle\nabla} )
- {}^E S(A^{+*}, \e, \e) \nonumber\\
&+& \tfrac{1}{4} (H(\r(\e),\r(\e))-R(\e,\e))(-,\r(A^{+*})) )\nonumber\\
&+& \tfrac{1}{2} (\psi^+, (B^+, (\nabla^{LC} \rho(\e), \widetilde{\psi}^{\scriptscriptstyle\nabla}) 
+ \tfrac{1}{3} \nabla {}^E S(\e,\e,\e))) \nonumber\\
&+& \tfrac{1}{2} (B^+, (\psi^+, (\nabla^{LC} \rho(\e), \widetilde{\psi}^{\scriptscriptstyle\nabla}) 
+ \tfrac{1}{3} \nabla {}^E S(\e,\e,\e))) \nonumber\\
&+& \tfrac{1}{2} (B^+, (B^+, (\nabla^{LC} \rho(\e), \psi^{\scriptscriptstyle\nabla}) 
+ (-\nabla^{LC} \rho(A), \widetilde{\psi}^{\scriptscriptstyle\nabla}) 
-\nabla {}^E S(A,\e,\e) \nonumber\\
&+& \tfrac{2}{3}R^{LC}(\widetilde{\psi}^{\scriptscriptstyle\nabla}, -, -, F)
- \tfrac{1}{6}(\nabla H(\r(\e), \r(\e)) -\nabla R(\e, \e))(-,F)))) \bigr\}\,,\\
S_3&=& \int \bigl\{ -\tfrac{1}{6} 
(B^+, (B^+, (B^+, \nabla \nabla^{LC} \rho(\e), \widetilde{\psi}^{\scriptscriptstyle\nabla})) 
+ \tfrac{1}{3} \nabla \nabla {}^E S(\e,\e,\e) )))  \nonumber\\
&-& (B^+, (B^+, (\tfrac{1}{4} \nabla^{LC} \rho(\e), (H(\r(\e), \r(\e)) -R(\e, \e))(-,B^+)))) \bigr\}\,.
\eea
$\epsilon^a$, $\psi_i$, and $\widetilde{\psi}$ are ghost fields.
$(\cdot, \cdot)$ is the pairing between $TM$ and $T^*M$, and $\langle \cdot, \cdot \rangle$ is the pairing between $E$ and $E^*$.
$R$ is the curvature tensor for the connection $\omega$, and $R^{LC}$ is the Riemann curvature tensor for the Levi-Civita connection $\mathring{\Gamma}$.
$\nabla^{LC} v := \nabla \nabla v - (R^{LC}, v)$ is the second order differential operator acting on a vector field $v \in \mathfrak{X}(M)$.


\section{Outlook}
We have discussed construction of a classical BV action functional based on 
underlining geometry.
Concrete constructions for higher $n \geq 3$ are unknown.
However, the $E$-torsion and the basic curvature have 
important roles.
QP-manifolds of mapping spaces, which appear in BV formalisms 
of general gauge theories have more complicated structures. 
They have many analyses but still are not well understood yet.

We have not discussed the quantization. To analyze relations of 
the quantization and geometry are an important problem.


\subsection*{Acknowledgment}
The author is grateful for support, hospitality and useful discussion within the workshop ``XLII Workshop on Geometric Methods in Physics.''

This work was supported by JSPS Grants-in-Aid for Scientific Research Number 22K03323.

\bibliographystyle{spmpsci}

\end{document}